\documentclass[leqno,11pt]{article}
\usepackage{amsfonts,latexsym}
\usepackage{amsmath}
\usepackage{amscd}
\usepackage{float,amsmath,amssymb,mathrsfs,bm,multirow,graphics}
\usepackage[dvips]{graphicx}

\newcommand{\nequation}{\setcounter{equation}{0}}
\renewcommand{\theequation}{\mbox{\arabic{section}.\arabic{equation}}}
\newcommand{\R}{{\Bbb R}}

\newcommand{\Z}{{\Bbb Z}}
\newcommand{\Q}{{\Bbb Q}}
\newcommand{\proofbegin}{\noindent{\it Proof.\,\,}}
\newcommand{\proofend}{\hfill$\Box$\bigskip}

\newtheorem{theorem}{Theorem}[section]
\newtheorem{proposition}[theorem]{Proposition}
\newtheorem{lemma}[theorem]{Lemma}

\newtheorem{remark}[theorem]{Remark}

\newtheorem{figuretext}{Figure}

\input epsf
\title{\sc On a novel integrable generalization of the nonlinear Schr\"odinger equation}
\author{J. Lenells and A.S. Fokas}
\date{{\small Department of Applied Mathematics and Theoretical Physics, University of Cambridge, Cambridge CB3 0WA, United Kingdom}}

\begin{document}
\maketitle
\begin{abstract} 

\noindent
We consider an integrable generalization of the nonlinear Schr\"odinger (NLS) equation that was recently derived by one of the authors using bi-Hamiltonian methods. This equation is related to the NLS equation in the same way that the Camassa Holm equation is related to the KdV equation. In this paper we: (a) Use the bi-Hamiltonian structure to write down the first few conservation laws. (b) Derive a Lax pair. (c) Use the Lax pair to solve the initial value problem. (d) Analyze solitons.
 \end{abstract}



\section{Introduction} \nequation
The celebrated nonlinear Sch\"odinger (NLS) equation appears in a wide range of physical circumstances from water waves to nonlinear optics. This is a consequence of the fact that it can be derived using physically meaningful asymptotic techniques from large classes of nonlinear PDE's \cite{Calogero}. Indeed, the NLS equation describes the amplitude $A(\xi, \tau)$ ($\xi = \epsilon x, \tau = \epsilon^2 t$) of the modulation of a monochromatic wave of a large class of dispersive nonlinear equations.

NLS belongs to a distinctive class of nonlinear equations called integrable. Integrable equations possess several special features including the existence of a Lax pair, as well as a bi-Hamiltonian formulation. Actually, the latter feature provides a simple algorithmic approach of constructing integrable equations. Indeed, the following result was derived independently in \cite{F-F1} and \cite{G-D1, G-D2}: Suppose that the operator $\theta_1 + \lambda \theta_2$ is Hamiltonian for all values of the parameter $\lambda$ (i.e. this operator is skew-symmetric and it satisfies an appropriate Jacobi identity). Then, each member of the hierarchy of equations
\begin{equation}\label{qtsum}
  q_t = \sum_{j = 1}^N c_j(\theta_2 \theta_1^{-1})^j q_x, \qquad n \in \Z, \quad c_j \; \text{constant},
\end{equation}
possesses infinitely many conserved quantities. Equation (\ref{qtsum}) with $c_1 = -1$, $N =1$, and 
\begin{equation}\label{kdvoperators}
\theta_1 = \partial, \qquad \theta_2 = \partial + \beta \partial^3 + \frac{1}{3}\alpha(q \partial + \partial q), \qquad \alpha, \beta \;\text{constants},
\end{equation}
yields the KdV equation 
$$q_t + q_x + \beta q_{xxx} + \alpha qq_x = 0.$$
Similarly, equation (\ref{qtsum}) with $c_1 = -1$, $N=1$, and 
$$\theta_1 = \partial + \nu \partial^3, \quad \theta_2 \; \text{as in (\ref{kdvoperators})}, \quad \nu \;\text{constant},$$
yields, with $q= u +\nu u_{xx}$, the Camassa-Holm equation
\begin{equation}\label{CH}
  u_t + u_x + \nu u_{txx} + \beta u_{xxx} + \alpha uu_x + \frac{1}{3} \alpha \nu (uu_{xxx} + 2u_xu_{xx}) = 0.
\end{equation}

The KdV equation provides an approximation to two-dimensional inviscid, irrotational water waves above a horizontal bottom $y = -h_0$ under the assumptions that the parameters $\alpha = a/h_0$ and $\beta = h_0^2/l^2$ are sufficiently small, where $a$ and $l$ are typical values of the amplitude and the wavelength of the waves. In these circumstances, one obtains the KdV if $O(\alpha) = O(\beta)$ and one neglects terms of $O(\alpha^2)$. It is shown in \cite{F-Liu} that if one retains terms of $O(\alpha^2)$ (but neglects terms of $O(\alpha^3)$), one obtains the Camassa-Holm equation. Although equation (\ref{CH}) was introduced by Fuchssteiner and one of the authors in 1981 \cite{F-F}, it began to attract a lot of interest (see for example \cite{ACHM, C-E0, C-M, HMPZ, L3, M}) only after its rederivation by Camassa and Holm \cite{C-H}. 

It was noted earlier that it is possible to derive mathematically the CH equation by utilizing the two Hamiltonian operators associated with the KdV equation. Similarly, by utilizing the two Hamiltonian operators associated with the NLS equation, it is possible to derive \cite{F} the following generalization of the NLS equation
\begin{equation}\label{GNLS}  
  iu_t - \nu u_{tx} + \gamma u_{xx} + \sigma |u|^2(u  + i \nu u_x) = 0, \qquad \sigma = \pm 1,
\end{equation}
depending on two real parameters $\gamma$ and $\nu$. In the same way that the CH equation provides a better approximation of water waves, equation (\ref{GNLS}) appears in the same circumstances as NLS provided that one retains terms of the next asymptotic order \cite{F}.
In this paper we: (a) Review the derivation of (\ref{GNLS}) and write down the first few conservation laws using the bi-Hamiltonian structure. (b) Derive a Lax pair of equation (\ref{GNLS}). (c) Use the Lax pair to solve the initial value problem. (d) Analyze solitons.

\section{Derivation and bi-Hamiltonian structure}\nequation
Consider the NLS equation
\begin{equation}\label{NLS}  
  iq_t + \gamma q_{xx} + \sigma q |q|^2 = 0, \qquad \sigma = \pm 1,
\end{equation}
where $\gamma$ is a real parameter. This equation is a particular reduction of the following bi-Hamiltonian system: Define the compatible pair of Hamiltonian operators
$$\sigma_3 = \begin{pmatrix} 1	& 0	\\
0	&	-1 \end{pmatrix} \quad \text{and} \quad \theta_2 = \begin{pmatrix} \gamma \partial_x + q \partial_x^{-1}r	& -q \partial_x^{-1}q	\\
-r \partial_x^{-1}r	&	\gamma\partial_x + r \partial_x^{-1}q \end{pmatrix};$$
the system
$$\begin{pmatrix} q	\\
r	\end{pmatrix}_t = i\theta_2 \sigma_3 \begin{pmatrix} q	\\
r	\end{pmatrix}_x,$$
reduces to equation (\ref{NLS}) when $r = \sigma \bar{q}$. If we replace $\sigma_3$ by $\theta_1 = \sigma_3 + i\nu I\partial_x$ where $I$ is the $2\times 2$ identity matrix, then the equation
$$\begin{pmatrix} q	\\
r	\end{pmatrix}_t = i\theta_2 \theta_1^{-1} \begin{pmatrix} q	\\
r	\end{pmatrix}_x,$$
together with the definition $q = u + i\nu u_x$, gives (\ref{GNLS}). 

Now let
$$\sigma_1 = \begin{pmatrix}
0 & 1 \\
1 & 0 \end{pmatrix},$$
and define two Poisson brackets for the functionals $F[q,r]$ and $G[q,r]$ by
$$\{F, G\}_\mathcal{D} = \int (\text{grad}\, F)^T \mathcal{D} \text{grad}\, G dx,$$
$$\{F, G\}_\mathcal{E} = \int (\text{grad}\, F)^T \mathcal{E} \text{grad}\, G dx,$$
where the superscript $T$ denotes the transpose and the operators $\mathcal{D}$ and $\mathcal{E}$ are defined by
$$\mathcal{D} = \theta_2\sigma_1 =\begin{pmatrix} 
 - q \partial_x^{-1}q & \gamma \partial_x + q \partial_x^{-1} r 
 \\  \gamma \partial_x + r \partial_x^{-1} q  & - r \partial_x^{-1} r \end{pmatrix}$$
and 
$$\mathcal{E} = -i\theta_1\sigma_1 = 
\begin{pmatrix}  0 & -i + \nu \partial_x 
 \\ i  + \nu \partial_x  & 0  \end{pmatrix}.$$
The gradient of a functional $F[q,r]$ is defined by
$$\text{grad}\, F = \begin{pmatrix} \frac{\delta F}{\delta q} \\ \frac{\delta F}{\delta r} \end{pmatrix},$$
provided that there exist functions $\frac{\delta F}{\delta q}$ and $\frac{\delta F}{\delta r}$ such that
$$\frac{d}{d\epsilon} F[q + \epsilon \varphi_1, r + \epsilon \varphi_2]\biggr|_{\epsilon =0} = \int \biggl(\frac{\delta F}{\delta q} \varphi_1 + \frac{\delta F}{\delta r} \varphi_2 \biggr)dx,$$
for any smooth functions $\varphi_1$ and $\varphi_2$.
Integration by parts shows that the operators $\mathcal{D}$ and $\mathcal{E}$ are skew-symmetric with respect to the bracket
 \begin{equation}\label{bracket}
   \left \langle \begin{pmatrix} f_1 \\ f_2 \end{pmatrix},  \begin{pmatrix} g_1 \\ g_2 \end{pmatrix} \right\rangle = \int (f_1g_1 + f_2 g_2) dx.
\end{equation}
If we introduce $u$ and $v$ by the relations 
\begin{equation}\label{quvrrelations}
  q = u + i\nu u_x, \qquad r = v - i\nu v_x,
\end{equation} 
we find that equation (\ref{GNLS}) has the bi-Hamiltonian formulation
$$\begin{pmatrix} q \\ r \end{pmatrix}_t =  \mathcal{D} \text{grad}\, H_1 = \mathcal{E} \text{grad}\, H_2,$$
where the Hamiltonians $H_1$ and $H_2$ are given by
$$H_1 = i\int u_x rdx = - i\int v_x qdx$$
and
$$H_2 = -\int \biggl(\gamma v u_{xx} + \frac{1}{2} q u v^2\biggr)dx = -\int \biggl(\gamma u v_{xx} + \frac{1}{2} r u^2 v\biggr)dx.$$
Indeed, if we write equation (\ref{GNLS}) as the system
$$\begin{pmatrix} q \\ r \end{pmatrix}_t = i\begin{pmatrix} \gamma u_{xx} + q u v \\ -\gamma v_{xx} - r u v \end{pmatrix},$$
then using
$$\text{grad}\, H_1 = \begin{pmatrix} -iv_x \\ iu_x \end{pmatrix},$$
we immediately find the formulation in terms of $H_1$. Moreover, from (\ref{quvrrelations}) and the definition (\ref{bracket}) of the bracket, we infer that
$$\text{grad}\, H_2 = \begin{pmatrix} (1 - i\nu \partial_x)^{-1} \delta H_2 / \delta u \\ (1 + i\nu \partial_x)^{-1} \delta H_2 / \delta v \end{pmatrix}.$$ 
Hence, using
$$\begin{pmatrix} \frac{\delta H_2}{\delta u} \\ \frac{\delta H_2}{\delta v} \end{pmatrix} = \begin{pmatrix} -\gamma v_{xx} - ruv \\ -\gamma u_{xx} - quv \end{pmatrix},$$
we find the formulation in terms of $H_2$. A hierarchy of conserved quantities $H_n$ can be obtained by means of the recursive relations
$$\mathcal{D} \text{grad}\, H_n = \mathcal{E} \text{grad}\, H_{n+1}, \qquad n \in \Z.$$
We find that
$$\mathcal{D} \text{grad}\, H_0 = \mathcal{E} \text{grad}\, H_1 = \begin{pmatrix}  q_x
 \\ r_x \end{pmatrix}$$
is satisfied for
$$H_0 = \frac{1}{\gamma}\int rq dx.$$
The first few conservation laws and associated Hamiltonian equations when $\bar{q} = \sigma r$ are summarized in Table \ref{GNLStable}.

\begin{table}[htdp]
\caption{Conservation laws of equation (\ref{GNLS})}
\begin{center}
\small
\begin{tabular}{|c|c|c|}\label{GNLStable}
$n$ & $H_n$ 	& $q_t = \mathcal{E} \text{grad}\, H_n$ \\
\hline
$2$	&	$-\int \left(\gamma v u_{xx} + \frac{1}{2} q u v^2\right)dx$	&	 $q_t = i\gamma u_{xx} + i\sigma q |u|^2 $	\\
$1$	&	$i\int u_x r$	&	$q_t =  q_x$	\\
$0$ 	& 	$\frac{1}{\gamma}\int rq dx$		&	$q_t = -\frac{i}{\gamma}(1 + i\nu \partial_x)q$		\\
\end{tabular}
\end{center}
\end{table}%

\section{A Lax pair} \nequation
Replacing $u(x,t)$ by $u(-x,t)$ if necessary, we can assume that $\nu$ has the same sign as $\gamma$ in equation (\ref{GNLS}). Let $\alpha = \gamma/\nu > 0$ and $\beta = 1/\nu$. Then the transformation 
$$u \to \beta \sqrt{\alpha} e^{i\beta x} u, \qquad \sigma \to -\sigma,$$ 
converts (\ref{GNLS}) into
\begin{equation}\label{GNLSgauge}  
  u_{tx} + \alpha\beta^2 u - 2i\alpha \beta u_x - \alpha u_{xx} + \sigma i \alpha \beta^2 |u|^2u_x = 0, \qquad \sigma = \pm 1.
\end{equation}
Thus, without loss of generality, in the remainder of this paper we will concentrate on equation (\ref{GNLSgauge}). This equation is the condition of compatibility of the system
\begin{equation}\label{lax}
\begin{cases}
	& \psi_x + i\zeta^2 \sigma_3 \psi = \zeta U_x \psi,	\\
	& \psi_t + i\eta^2\sigma_3 \psi = \left[\alpha\zeta U_x +\frac{i\alpha\beta^2}{2}\sigma_3\left(\frac{1}{\zeta} U-U^2\right)\right]\psi,
\end{cases}
\end{equation}	
where $\psi(x,t)$ is a $2\times2$-matrix valued function and
\begin{equation}\label{Uetadef}
  U =\begin{pmatrix} 0	& u	\\
v	&	0 \end{pmatrix}, \qquad \eta = \sqrt{\alpha}\left(\zeta - \frac{\beta}{2\zeta}\right), \qquad v = \sigma \bar{u}.
\end{equation}
Notice that the $x$-part of (\ref{lax}) has the same form as the $x$-part of the Lax pair for the derivative NLS equation derived in the pioneering work of Kaup and Newell \cite{K-N}. The fact that the $x$-part of (\ref{lax}) contains only the $x$-derivative of $U$ is related to the absence of a $u_t$-term in the gauge transformed equation (\ref{GNLSgauge}).

\section{The solution of the Cauchy problem} \nequation
In this section we will use spectral analysis to express the solution of the Cauchy problem of (\ref{GNLSgauge}) in terms of the solution of an appropriate Riemann-Hilbert problem. 
Letting
\begin{equation}\label{psimurelation} 
  \psi(x,t,\zeta) = e^{\frac{i}{2}\int^{x}_{-\infty} v_x(x',t) u_x(x',t) dx'\hat{\sigma}_3}\mu(x, t, \zeta) D(x,t) e^{-i(\zeta^2x + \eta^2t)\sigma_3},
\end{equation}
where $\hat{\sigma}_3$ acts on a $2\times 2$ matrix $A$ by $\hat{\sigma}_3A = \sigma_3 A \sigma_3^{-1}$ and the diagonal matrix $D$ is defined by
\begin{equation}\label{Ddef}  
  D(x, t) = e^{-\frac{i}{2}\int_x^\infty v_x(x',t) u_x(x',t) dx' \sigma_3},
\end{equation}
the Lax pair (\ref{lax}) becomes
\begin{equation}\label{mulax}
\begin{cases}
	& \mu_x + i\zeta^2 [\sigma_3, \mu] = V_1\mu, \\
	& \mu_t + i\eta^2 [\sigma_3, \mu] = V_2\mu,
\end{cases}
\end{equation}
where
\begin{align}\label{V1explicit}
& V_1 = e^{-\frac{i}{2}\int^{x}_{-\infty} v_x u_xdx' \hat{\sigma}_3}\left(\zeta U_x - \frac{i}{2}v_xu_x \sigma_3\right),
	\\ \label{V2explicit}
& V_2 = e^{-\frac{i}{2}\int^{x}_{-\infty} v_x u_xdx' \hat{\sigma}_3}\left(\alpha\zeta U_x +\frac{i\alpha\beta^2}{2}\sigma_3\left(\frac{1}{\zeta} U-U^2\right) - i\frac{\alpha}{2}(u_xv_x - \beta^2 uv) \sigma_3 \right).
\end{align}
In order to derive (\ref{mulax}) from (\ref{lax}) we use the conservation law
\begin{equation}\label{conslaw}  
  (u_xv_x)_t = \alpha(u_xv_x - \beta^2 uv)_x
\end{equation}  
to carry out the differentiation of $D$ and of the exponential prefactor on the right-hand side of (\ref{psimurelation}) with respect to $t$.

Before using the spectral analysis of (\ref{mulax}) to derive a nonlinear Fourier transform pair, we note that the transformation (\ref{psimurelation}) can be motivated as follows: If $U_x$ tends to zero at infinity, then there exists a solution of (\ref{lax}) which tends to $\exp(-i(\zeta^2x + \eta^2t)\sigma_3)$. This suggests the transformation $\psi = \Psi\exp(-i(\zeta^2x + \eta^2t) \sigma_3)$ and then $\Psi$ solves
\begin{equation}\label{Psilax}  
\begin{cases}
	& \Psi_x + i\zeta^2[\sigma_3, \Psi] = \zeta U_x \Psi,	\\
	& \Psi_t + i\eta^2[\sigma_3, \Psi] = \left[\alpha\zeta U_x +\frac{i\alpha\beta^2}{2}\sigma_3\left(\frac{1}{\zeta} U-U^2\right)\right]\Psi.
\end{cases}
\end{equation}
We seek a solution of this equation in the form
$$\Psi = D + \frac{\Psi_1}{\zeta} + \frac{\Psi_2}{\zeta^2} + O\left(\frac{1}{\zeta^3}\right), \quad \zeta \to \infty,$$
where $D, \Psi_1, \Psi_2$ are independent of $\zeta$. Substituting the above expansion in the $x$-part of (\ref{Psilax}) it follows from the $O(\zeta^2)$ terms that $D$ is a diagonal matrix. Furthermore, one finds the following equations for the $O(\zeta)$ and the diagonal part of the $O(1)$ terms
\begin{equation}\label{Ozetaterms}  
  O(\zeta): i[\sigma_3, \Psi_1] = U_xD, \quad \text{i.e.} \quad \Psi_1^{(o)} = \frac{i}{2}U_xD\sigma_3,
\end{equation}
where $\Psi_1^{(o)}$ denotes the off-diagonal part of $\Psi_1$;
$$O(1): D_x = U_x\Psi_1^{(o)}, \quad \text{i.e.} \quad D_x = \frac{i}{2}u_x v_x \sigma_3 D.$$
This leads to the definition (\ref{Ddef}) of $D$. In fact, a similar analysis of the $t$-part of (\ref{Psilax}) reveals that $D_t = \frac{i \alpha}{2}(u_xv_x - \beta^2 uv)\sigma_3 D$, which, in view of the conservation law (\ref{conslaw}), is compatible with (\ref{Ddef}). Therefore, up to a multiplicative constant, the asymptotic behavior of $\Psi$ is given by $D$ as defined in (\ref{Ddef}). In order to formulate a Riemann-Hilbert problem for the solution of the inverse spectral problem, we seek solutions of the spectral problem which approach the $2 \times 2$ identity matrix $I$ as $\zeta \to \infty$. This suggests introducing $\mu$ by $\psi = \mu D e^{-i(\zeta^2x + \eta^2t)\sigma_3}$, so that
\begin{equation}\label{muasymptotic}  
  \mu = I + O\left(\frac{1}{\zeta}\right), \qquad \zeta \to \infty.
\end{equation}
The additional exponential factor in (\ref{psimurelation}), which does not affect (\ref{muasymptotic}), is convenient because it implies that two different solutions of (\ref{mulax}) are related by multiplication by a matrix which is independent of $U_x$ (see equation (\ref{mutildemu}) below).

\subsection{A nonlinear Fourier transform}
We now consider the spectral analysis of the $x$-part of (\ref{mulax}). Since the analysis will take place at a fixed time the $t$-dependence will be suppressed. Define two solutions $\mu_1$ and $\mu_2$ of the $x$-part of (\ref{mulax}) by
\begin{align} \label{mu1def}
  \mu_1(x,\zeta) = I + \int_{-\infty}^x e^{i \zeta^2 (x' - x) \hat{\sigma}_3} (V_1 \mu_1)(x',\zeta)dx',
 	\\ \label{mu2def}
  \mu_2(x,\zeta) = I - \int_x^{\infty} e^{i \zeta^2 (x' - x) \hat{\sigma}_3} (V_1 \mu_2)(x',\zeta)dx'.
\end{align}
The second columns of these matrix equations involve $\exp[2i\zeta^2(x' - x)]$. It follows that the second column vectors of $\mu_1$ and $\mu_2$ are bounded and analytic for $\zeta$ in $D_-$ and $D_+$, respectively, where 
$$D_+ = \{\zeta | \arg \zeta \in (0, \pi/2) \cup (\pi, 3\pi/2)\}$$ 
is the union of the first and third quadrants, while 
$$D_- = \{\zeta | \arg \zeta \in (\pi/2, \pi) \cup (3\pi/2, 2\pi)\}$$ 
is the union of the second and fourth quadrants. We will denote these vectors with superscripts $(-)$ and $(+)$ to indicate these boundedness properties.
Similar conditions are valid for the first column vectors, hence
$$\mu_1 = \left(\mu_1^{(+)}, \mu_1^{(-)}\right), \qquad 
\mu_2 = \left(\mu_2^{(-)}, \mu_2^{(+)}\right).$$ 
Note that $\mu(x,\zeta) = \mu_j(x,\zeta)$, $j = 1,2$, satisfies the symmetry relations
\begin{align}\label{musymmetries1}
  \mu_{11}(x,\zeta) = \overline{\mu_{22}(x,\sigma\bar{\zeta})}, \qquad \mu_{21}(x,\zeta) = \overline{\mu_{12}(x, \sigma\bar{\zeta})},
\end{align}
as well as
\begin{align}\label{musymmetries2}
  \mu_{11}(x, -\zeta) = \mu_{11}(x,\zeta), \qquad \mu_{12}(x, -\zeta) = -\mu_{12}(x, \zeta), 
  		\\\nonumber
    \mu_{21}(x, -\zeta) = -\mu_{21}(x, \zeta), \qquad \mu_{22}(x, -\zeta) = \mu_{22}(x, \zeta).
\end{align}

\subsubsection{The function $s(\zeta)$}
Any two solutions $\mu$ and $\tilde{\mu}$ of (\ref{mulax}) are related by an equation of the form
\begin{equation}\label{mutildemu}  
  \mu(x,\zeta) = \tilde{\mu}(x,\zeta)  e^{-i\zeta^2 x\hat{\sigma}_3} N_0(\zeta),
\end{equation}
where $N_0(\zeta)$ is a $2\times 2$ matrix independent of $x$. 
Indeed, let $\psi$ and $\tilde{\psi}$ be the solutions of 
\begin{equation}\label{psixpart}  
  \psi_x + i\zeta^2 \sigma_3 \psi = \zeta U_x \psi,
\end{equation}  
related to $\mu$ and $\tilde{\mu}$ as
$$\psi(x,\zeta) = e^{\frac{i}{2}\int^{x}_{-\infty} v_x u_x dx'\hat{\sigma}_3}\mu(x, \zeta) D e^{-i\zeta^2 x \sigma_3}.$$
Then, since the first and second columns of a solution to (\ref{psixpart}) satisfy the same equation, there exists a $2 \times 2$ matrix $N_1(\zeta)$ independent of $x$ such that
\begin{equation}\label{psirelations}  
  \psi(x, \zeta) = \tilde{\psi}(x, \zeta) N_1(\zeta).
\end{equation}
Comparing equations (\ref{mutildemu}) and (\ref{psirelations}) we find $$N_0(\zeta) = e^{-\frac{i}{2}\int_{-\infty}^\infty v_xu_x dx \hat{\sigma}_3} N_1(\zeta).$$
Thus we can define the spectral function $s(\zeta)$ by 
\begin{equation}\label{seq} 
  \mu_2(x,\zeta) = \mu_1(x,\zeta)e^{-i\zeta^2 x\hat{\sigma}_3} s(\zeta), \qquad \text{Im}\,\zeta^2 = 0.
\end{equation}
Evaluation at $x \to -\infty$ gives
\begin{equation}\label{sexplicit}
  s(\zeta) = I - \int_{-\infty}^\infty e^{i \zeta^2x \hat{\sigma}_3} (V_1 \mu_2)(x,\zeta)dx, \qquad \text{Im}\, \zeta^2 = 0.
\end{equation}
Moreover, it follows from (\ref{mulax}) that the determinant of $\mu$ is independent of $x$, so that evaluation of $\det \mu_j$, $j = 1,2$, at $x = \pm \infty$ shows that
\begin{equation}\label{detmuisone}  
  \det \mu_j = 1, \qquad j= 1,2.
\end{equation}
In particular,
$$\det s(\zeta) = 1.$$
From (\ref{musymmetries1}) and (\ref{musymmetries2}), we deduce that
$$s_{11}(\zeta) = \overline{s_{22}(\bar{\zeta})}, \qquad s_{21}(\zeta) = \sigma\overline{s_{12}(\bar{\zeta})}.$$
This analysis suggests the following notation for $s$:
$$s(\zeta) = \begin{pmatrix} \overline{a(\bar{\zeta})} & b(\zeta) \\
\sigma \overline{b(\bar{\zeta})} 	&	a(\zeta) \end{pmatrix}.$$
From the explicit expression (\ref{sexplicit}) for $s(\zeta)$ and the fact that the second column of $\mu_2$ is defined and analytic in $D_+$, we find that $a(\zeta)$ has an analytic continuation to all of $D_+$.
The symmetry (\ref{musymmetries2}) implies that $a(\zeta)$ is an even function of $\zeta$, whereas $b(\zeta)$ is an odd function of $\zeta$.

\subsubsection{Residue conditions}
Since $a(\zeta)$ is an even function, each zero $\zeta_j$ of $a(\zeta)$ is accompanied by another zero at $-\zeta_j$. We assume that $a(\zeta)$ has $2N$ simple zeros $\{\zeta_j\}_{j = 1}^{2N} \subset D_+$ such that $\{\zeta_j\}_{j = 1}^{N}$ belong to the first quadrant and $\zeta_{j+N} = -\zeta_j$, $j = 1, \dots, N$.

The second column of equation (\ref{seq}) is
\begin{equation}\label{mu2amu1b}
  \mu_2^{(+)} = a\mu_1^{(-)} + b\mu_1^{(+)}e^{-2i\zeta^2 x}, \qquad \zeta \in \R \cup i\R.
\end{equation}
Applying $\det \left( \mu_1^{(+)}, \cdot \right)$ to (\ref{mu2amu1b}) and recalling (\ref{detmuisone}), we find
$$\det \left( \mu_1^{(+)}(x, \zeta), \mu_2^{(+)}(x, \zeta) \right) = a(\zeta), \qquad \zeta \in D_+,$$
where we have used that both sides are well-defined and analytic in $D_+$ to extend the above relation to all of $D_+$.
Hence, if $a(\zeta_j) = 0$, then $\mu_1^{(+)}(x, \zeta_j)$ and $\mu_2^{(+)} (x, \zeta_j)$ are linearly dependent vectors for each $x$. It follows that the $2 \times 2$ matrix 
$$\left(\psi_1^{(+)} (x, \zeta_j), \psi_2^{(+)} (x, \zeta_j) \right) = 
e^{\frac{i}{2}\int_{-\infty}^x v_x u_x dx' \hat{\sigma}_3} \left( \mu_1^{(+)}(x, \zeta_j), \mu_2^{(+)} (x, \zeta_j) \right) D e^{-i\zeta_j^2 x \sigma_3},$$
 is a solution of the $x$-part of (\ref{lax}) with linearly dependent column vectors. We conclude that there exists a constant $b_j$ such that
$$\psi_1^{(+)}(x, \zeta_j) = b_j \psi_2^{(+)}(x, \zeta_j), \qquad x \in \R.$$
This implies
\begin{equation}\label{mubj}  
  \mu_1^{(+)}(x, \zeta_j) = b_j e^{i\int_{-\infty}^\infty v_x u_x dx}e^{2i\zeta_j^2 x} \mu_2^{(+)}(x, \zeta_j), \qquad x \in \R.\end{equation}
Recalling the symmetries (\ref{musymmetries1})-(\ref{musymmetries2}), the complex conjugate of (\ref{mubj}) is
$$\mu_1^{(-)}(x, \bar{\zeta}_j) = \bar{b}_j e^{-i\int_{-\infty}^\infty v_x u_x dx}e^{-2i\bar{\zeta}_j^2 x} \mu_2^{(-)}(x, \bar{\zeta}_j), \qquad x \in \R.$$
Consequently, the residues of $\mu_1^{(+)}(x,\zeta)/a(\zeta)$ and $\mu_1^{(-)}(x,\zeta)/\overline{a(\bar{\zeta})}$ at $\zeta_j$ and $\bar{\zeta}_j$ are
$$\underset{\zeta_j}{\text{Res}} \frac{\mu_1^{(+)}(x, \zeta)}{a(\zeta)} = \frac{\mu_1^{(+)}(x, \zeta_j)}{\dot{a}(\zeta_j)} =  C_j e^{2i\zeta_j^2 x} \mu_2^{(+)}(x, \zeta_j),$$
$$\underset{\bar{\zeta}_j}{\text{Res}}  \frac{\mu_1^{(-)}(x, \zeta)}{\overline{a(\bar{\zeta})}} = \frac{\mu_1^{(-)}(x, \bar{\zeta}_j)}{\overline{\dot{a}(\zeta_j)}} =  \bar{C}_je^{-2i\bar{\zeta}_j^2 x} \mu_2^{(-)}(x, \bar{\zeta}_j),$$
where $\dot{a}(\zeta) = \frac{da}{d\zeta}$ and
$$C_j = \frac{b_j e^{i\int_{-\infty}^\infty v_x u_x dx}}{\dot{a}(\zeta_j)}.$$

In fact, there are only $N$ independent constants $C_j$. Indeed, using the symmetries (\ref{musymmetries2}) it follows from (\ref{mubj}) that $b_{j + N} = -b_j$, $j = 1, \dots, N.$ The fact that $\dot{a}$ is an odd function implies the relations $C_{j + N} = C_j$, $j = 1, \dots, N.$

\subsubsection{The inverse problem}\label{inverseproblemsubsec}
The inverse problem involves reconstructing $U$ from the spectral functions $\mu_j(x,\zeta)$, $j = 1,2$. We obtained in (\ref{Ozetaterms}) that $\Psi_1^{(o)} = \frac{i}{2}U_xD\sigma_3$ whenever 
$$\Psi = D + \frac{\Psi_1}{\zeta} + \frac{\Psi_2}{\zeta^2} + O\left(\frac{1}{\zeta^3}\right), \qquad \zeta \to \infty,$$
is a solution of (\ref{Psilax}). 
This implies that
\begin{equation}\label{recoverq}  
  u_x(x) = 2im(x)e^{i\int_{-\infty}^{x} v_xu_x dx'},
\end{equation}
where 
$$\mu = I + \frac{m^{(1)}}{\zeta} + \frac{m^{(2)}}{\zeta^2} + O\left(\frac{1}{\zeta^3}\right), \qquad \zeta \to \infty,$$
is the corresponding solution of (\ref{mulax}) and we write $m(x)$ for $m^{(1)}_{12}(x)$. From equation (\ref{recoverq}) and its complex conjugate, we obtain $u_xv_x = 4 \sigma |m|^2.$ Thus, $u_x$ is expressed in terms of $m$ by
\begin{equation}\label{recoverux}
  u_x(x) = 2im(x)e^{4 \sigma i\int_{-\infty}^{x} |m|^2 dx'}.
 \end{equation}

\subsubsection{The Riemann-Hilbert problem}
\begin{figure}
\begin{center}
   \includegraphics[width=.4\textwidth]{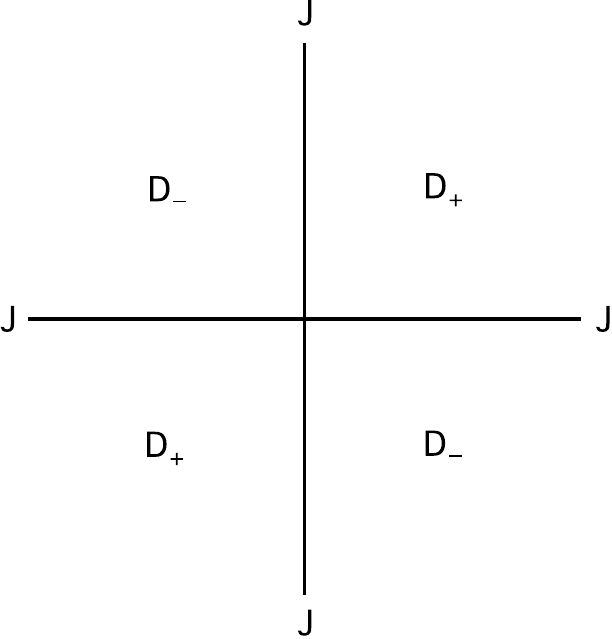} \\
     \begin{figuretext}\label{GNLSRiemannHilbert.pdf}
        Illustration of the contour and the four sections of the Riemann-Hilbert problem (\ref{MplusMminusdef}) in the complex $\zeta$-plane.
     \end{figuretext}
     \end{center}
\end{figure}

Equation (\ref{seq}) can be rewritten in a form expressing the jump condition of a $2 \times 2$ RH problem. Algebraic manipulations give
$$M_-(x,\zeta) = M_+(x, \zeta)J(x,\zeta), \qquad \zeta \in \R \cup i\R,$$
where the matrices $M_-$, $M_+$, $J$ are defined by
\begin{align}\label{MplusMminusdef}
M_+ = \left(\frac{\mu_1^{(+)}}{a(\zeta)}, \mu_2^{(+)}\right), \qquad \zeta \in \bar{D}_+; \qquad
M_- = \left(\mu_2^{(-)}, \frac{\mu_1^{(-)}}{\overline{a(\bar{\zeta})}}\right), \qquad \zeta \in \bar{D}_-;
\end{align}
\begin{equation}\label{Jdef}
J(x,\zeta) = \begin{pmatrix} 1	&	-\frac{b(\zeta)}{\overline{a(\bar{\zeta})}}e^{-2i\zeta^2 x} 	\\
\sigma \frac{\overline{b(\bar{\zeta})}}{a(\zeta)} e^{2i\zeta^2 x}	&	\frac{1}{a(\zeta)\overline{a(\bar{\zeta})}} \end{pmatrix}, \qquad \zeta \in \R \cup i\R.
\end{equation}
The contour for this RH problem is depicted in Figure \ref{GNLSRiemannHilbert.pdf}. We summarize our discussion in the following theorem, in which we use the notation $[A]_1$ ($[A]_2$) for the first (second) column of a $2\times 2$ matrix $A$. 

\begin{theorem}\label{RHtheorem}
\begin{enumerate}
\item Given $u_x(x)$ with sufficient smoothness and decay, define the map
$$\mathbb{S}:\{u_x(x)\} \to \{a(\zeta), b(\zeta), C_j\}$$
by the following equations:
\begin{equation}\label{abRH}
  \begin{pmatrix} b(\zeta) \\
 a(\zeta) \end{pmatrix} = [s(\zeta)]_2, \qquad s(\zeta) = \lim_{x \to -\infty} e^{i\zeta^2 x \hat{\sigma}_3}\mu_2(x,\zeta), \qquad \zeta \in \R \cup i\R,
 \end{equation}
and
\begin{equation}\label{CjRH}
  [\mu_2(x,\zeta_j)]_{2}\dot{a}(\zeta_j) C_j = [\mu_1(x,\zeta_j)]_{1}e^{-2i\zeta_j^2 x},\qquad j = 1, \dots, 2N.
 \end{equation}
The functions $\mu_1(x,\zeta)$ and $\mu_2(x,\zeta)$ are the unique solutions of the Volterra linear integral equations (\ref{mu1def})-(\ref{mu2def}), and we assume that $a(\zeta)$ has $2N$ simple zeros $\{\zeta_j\}_{j = 1}^{2N} \subset D_+$ ordered so that $\{\zeta_j\}_{j = 1}^{N}$ belong to the first quadrant and $\zeta_{j+N} = -\zeta_j$, $j = 1, \dots, N$.

The spectral data $\{a(\zeta), b(\zeta), C_j\}$ have the following properties:
\begin{itemize}
\item $a(\zeta)$ is defined for $\zeta \in \bar{D}_+$ and analytic in $D_+$.
\item $b(\zeta)$ is defined for $\text{\upshape Im}\,\zeta^2 = 0$.
\item $a(\zeta)\overline{a(\bar{\zeta})} - \sigma b(\zeta)\overline{b(\bar{\zeta})} = 1, \qquad \text{\upshape  Im}\, \zeta^2 = 0$.
\item $a(\zeta) = 1 + \left(\frac{1}{\zeta}\right), \qquad \zeta \to \infty, \quad \zeta \in \bar{D}_+.$
\item $b(\zeta) = \left(\frac{1}{\zeta}\right), \qquad \zeta \to \infty, \quad \text{\upshape  Im}\, \zeta^2 = 0.$
\item $a(\zeta)$ and $b(\zeta)$ satisfy the symmetry relations
\begin{equation}\label{abevenoddsymmetries}  
  a(-\zeta) = a(\zeta), \quad \zeta \in \bar{D}_+, \qquad b(-\zeta) = -b(\zeta), \quad  \text{\upshape Im}\, \zeta^2 = 0.
\end{equation}
\item $C_{j + N} = C_j, \qquad j = 1, \dots, N.$
\end{itemize}

\item The map $\Q:\{a(\zeta), b(\zeta), C_j\} \mapsto \{u_x(x)\}$, inverse to $\mathbb{S}$, is defined by
\begin{equation}\label{recoverqRH}
u_x(x) = 2im(x)e^{4 \sigma i\int_{-\infty}^{x} |m|^2 dx'}, \qquad m(x) = \lim_{\zeta \to \infty} (\zeta M(x, \zeta))_{12},
\end{equation}
where $M(x,\zeta)$ is the unique solution of the following RH problem:
\begin{itemize}
\item $M(x,\zeta) = \left\{ \begin{array}{ll}
M_+(x,\zeta) &  \zeta \in D_+, \\
M_-(x,\zeta) &  \zeta \in D_-, \\
\end{array} \right.$ 

is a sectionally meromorphic function.

\item $M_-(x,\zeta) = M_+(x, \zeta)J(x,\zeta)$ for $\zeta \in \R \cup i\R,$ where $J$ is defined in (\ref{Jdef}).

\item $M(x,\zeta)$ has the asymptotic behavior
\begin{equation}\label{MtoI}
M(x,\zeta) = I + O\left(\frac{1}{\zeta}\right), \qquad \zeta \to \infty.
\end{equation}

\item The first column of $M_+$ has simple poles at $\zeta = \zeta_j$, $j = 1, \dots, 2N$, and the second column of $M_-$ has simple poles at $\zeta = \bar{\zeta}_j$, $j = 1, \dots, 2N$.
The associated residues are given by
\begin{align}\label{residue1}
\underset{\zeta_j}{\text{\upshape Res}} [M(x,\zeta)]_1 =& C_j e^{2i\zeta_j^2 x} [M(x,\zeta_j)]_2, \qquad j = 1, \dots, 2N,
		\\\label{residue2}
\underset{\bar{\zeta}_j}{\text{\upshape Res}} [M(x,\zeta)]_2 =& \bar{C}_j e^{-2i\bar{\zeta}_j^2 x} [M(x,\bar{\zeta}_j)]_1, \qquad j = 1, \dots, 2N.
\end{align}

\end{itemize}

\item We have
$$\mathbb{S}^{-1} = \Q.$$

\end{enumerate}
\end{theorem}

In the case when $a(\zeta)$ has no zeros, the unique solvability of the Riemann-Hilbert problem in Theorem \ref{RHtheorem} is a consequence of the following vanishing lemma, while if $a(\zeta)$ has zeros the singular RH problem can be mapped to a regular one following the approach of \cite{F-I}.

\begin{lemma}[Vanishing lemma]
The Riemann-Hilbert problem in Theorem \ref{RHtheorem} with the vanishing boundary condition
$$M(x, \zeta) \to 0 \quad \text{as} \quad \zeta \to \infty,$$
has only the zero solution.
\end{lemma}
\proofbegin
Assume that $M(x,\zeta)$ is a solution of the RH problem in Theorem \ref{RHtheorem} such that $M_\pm(x,\zeta) \to 0$ as $\zeta \to \infty$.
Let $A^\dagger$ denote the complex conjugate transpose of a matrix $A$. Define
\begin{align*}
& H_+(\zeta) = M_+ (\zeta)M_-^{\dagger}(-\sigma\bar{\zeta}), \qquad \zeta \in D_+, 
	\\
& H_-(\zeta) = M_-(\zeta)M_+^{\dagger}(-\sigma\bar{\zeta}),  \qquad \zeta \in D_-,
\end{align*}
where we have suppressed the $x$ dependence. The functions
$H_+(\zeta)$ and $H_-(\zeta)$ are analytic in the interior of $D_+$ and $D_-$, respectively. By the symmetry relations (\ref{abevenoddsymmetries}) we infer that
$$J^\dagger(-\sigma \bar{\zeta}) = J(\zeta), \qquad \zeta \in \R \cup i\R.$$
Since
$$H_+(\zeta) = M_+ (\zeta)J^{\dagger}(-\sigma\bar{\zeta})M_+^{\dagger}(-\sigma\bar{\zeta}), \quad H_-(\zeta) = M_+(\zeta)J(\zeta)M_+^{\dagger}(-\sigma\bar{\zeta}), \quad \zeta \in \R \cup i\R,$$
it follows that $H_+(\zeta) = H_-(\zeta)$ on $\R \cup i\R$. Therefore, $H_+(\zeta)$ and $H_-(\zeta)$ define an entire function vanishing at infinity, and so $H_+(\zeta)$ and $H_-(\zeta)$ are identically zero.

Let us first consider the case $\sigma = 1$. In this case $J(is)$ is a hermitian matrix with unit determinant and $(11)$ entry $1$ for any $s$. Hence, $J(is)$ is a positive definite matrix. Since $H_-(is)$ vanishes identically for $s \in \R$, we find
$$M_+(is)J(is)M_+^{\dagger}(is) = 0,\qquad s\in\R,$$
and so $M_+(is) = 0$ for $s \in \R$. By analytic continuation, $M_+$ and $M_-$ vanish identically. This proves the lemma when $\sigma = 1$.

If $\sigma = -1$, we note that $J(s)$ is a hermitian positive definite matrix for any $s \in \R$. Since $H_-(s)$ vanishes identically for $s \in \R$, we find
$$M_+(s)J(s)M_+^{\dagger}(s) = 0, \qquad s \in \R,$$
and so $M_+(s) = 0$ for $s \in \R$. Again it follows that $M_+$ and $M_-$ vanish identically, which proves the lemma when $\sigma = -1$.
\proofend

\begin{remark}\upshape
It follows from the symmetries (\ref{musymmetries1})-(\ref{musymmetries2}) that the solution $M(x, \zeta)$ of the Riemann-Hilbert problem in Theorem \ref{RHtheorem} respects the symmetries
\begin{align} \nonumber
& M_{11}(x,\zeta) = \overline{M_{22}(x, \sigma \bar{\zeta})}, \qquad M_{21}(x,\zeta) = \overline{M_{12}(x, \sigma \bar{\zeta})},
	\\  \label{Msymmetry}
 & M_{11}(x, - \zeta) = M_{11}(x,\zeta), \qquad M_{12}(x, - \zeta) = - M_{12}(x,\zeta).
	\\ \nonumber
 & M_{21}(x, - \zeta) = - M_{21}(x,\zeta), \qquad M_{22}(x, - \zeta) = M_{22}(x,\zeta).
\end{align}
Moreover, if one enforces (\ref{Msymmetry}), then only one of the two residue conditions (\ref{residue1})-(\ref{residue2}) needs to be verified since the other condition is a consequence of symmetry.
\end{remark}

\subsection{Time evolution}
Suppose $u(x,t)$ is a solution of equation (\ref{GNLSgauge}). At each time $t$, we define the spectral data $a(\zeta,t)$, $b(\zeta,t)$, and $\{C_j(t)\}_{j=1}^{2N}$ according to (\ref{abRH}) and (\ref{CjRH}), i.e.
\begin{equation}\label{abtime}
  \begin{pmatrix} b(\zeta,t) \\
 a(\zeta,t) \end{pmatrix} = [s(\zeta,t)]_2, \qquad s(\zeta,t) = \lim_{x \to -\infty} e^{i\zeta^2 x \hat{\sigma}_3}\mu_2(x,t,\zeta), \qquad \zeta \in \R \cup i\R,
 \end{equation}
and
\begin{equation}\label{Cjtime}
  [\mu_2(x,t, \zeta_j(t))]_{2}\dot{a}(\zeta_j(t), t) C_j(t) = [\mu_1(x,t,\zeta_j(t))]_{1}e^{-2i\zeta_j^2(t) x},\qquad j = 1, \dots, 2N.
 \end{equation}
\begin{proposition}\label{timeevolutionprop}
  The evolution of the spectral data is given by
\begin{equation}\label{abevolution}
  a(\zeta, t) = a(\zeta, 0), \quad b(\zeta, t) = b(\zeta, 0)e^{-2i\eta^2 t}, \quad \zeta_j(t) = \zeta_j(0), \quad C_j(t) = C_j(0)e^{2i\eta_j^2 t},
  \end{equation}
where $\eta$ was defined in (\ref{Uetadef}) and $\eta_j = \sqrt{\alpha}\left(\zeta_j - \frac{\beta}{2\zeta_j}\right)$.
\end{proposition} 
\proofbegin
From the $t$-part of (\ref{mulax}), we infer that
\begin{equation}\label{mu2t}
  \left(e^{i(\zeta^2 x + \eta^2 t) \hat{\sigma}_3}\mu_2(x,t,\zeta)\right)_t = e^{i(\zeta^2 x + \eta^2 t) \hat{\sigma}_3}(V_2\mu_2)(x,t,\zeta).
\end{equation}
Assuming that $u$ has sufficient decay at infinity, we have $V_2 \to 0$ as $x\to \pm \infty$. Evaluation of (\ref{mu2t}) at $x \to -\infty$ gives
$$\left(e^{i\eta^2 t \hat{\sigma}_3}s(\zeta,t)\right)_t = 0, \qquad \zeta \in \R \cup i\R.$$
The second column of this equation gives the time evolution of $a$ and $b$ as stated in (\ref{abevolution}).
Since $a_t \equiv 0$, the eigenvalues $\zeta_j$ are time-independent. Moreover, differentiating the first row of (\ref{Cjtime}) with respect to $t$ and evaluating the resulting equation at $x \to -\infty$, we find
$$C_{jt} = 2i\eta^2C_j,\qquad j = 1, \dots, 2N.$$
This completes the proof. 
\proofend

\subsection{The Cauchy problem}
The Cauchy problem for equation (\ref{GNLS}) with initial data $u_{0}(x)$ with sufficient smoothness and decay can now be solved using only linear operations as follows:

\begin{enumerate}
\item Define $\mu_1(x,\zeta)$ and $\mu_2(x,\zeta)$ in terms of $u_{0x}$ as the unique solutions of the Volterra linear integral equations (\ref{mu1def})-(\ref{mu2def}), and use them to obtain the spectral data $\{a(\zeta,0), b(\zeta,0), C_j(0)\}$ at time $t = 0$ according to (\ref{abRH})-(\ref{CjRH}).
\item Obtain the spectral data $\{a(\zeta,t), b(\zeta,t), C_j(t)\}$ at a later time $t$ from equation (\ref{abevolution}).
\item Solve the RH problem in Theorem \ref{RHtheorem} with the jump matrix $J(x,t,\zeta)$ defined in terms of $a(\zeta,t)$ and $b(\zeta,t)$ by (\ref{Jdef}) and with the residue conditions (\ref{residue1})-(\ref{residue2}) defined in terms of $C_j(t)$.
\item Use the solution $M(x,t,\zeta)$ to obtain $u_x(x,t)$ according to (\ref{recoverqRH}).
\item Obtain $u(x,t)$ by integration.
\end{enumerate}

Observe that (\ref{GNLSgauge}) is an evolution equation in $u_x$ and that any solution $u(x,t)$ is undetermined up to $u(x,t) \to u(x,t) + h(t)$ for an arbitrary function $h(t)$. The requirement that $u$ goes to zero as $|x| \to \infty$ removes this non-uniqueness.

\section{Solitons} \nequation
The solitons correspond to spectral data $\{a(\zeta), b(\zeta), C_j\}$ for which $b(\zeta)$ vanishes identically. In this case the jump matrix $J$ in (\ref{Jdef}) is the identity matrix and the Riemann-Hilbert problem of Theorem \ref{RHtheorem} consists of finding a meromorphic function $M(x,\zeta)$ satisfying (\ref{MtoI}) as well as the residue conditions (\ref{residue1})-(\ref{residue2}).
From (\ref{MtoI}) and (\ref{residue1}) we get
\begin{equation}\label{M1decompose}
[M(x,\zeta)]_1 = \begin{pmatrix} 1	\\	0 \end{pmatrix} + \sum_{j = 1}^{2N} \frac{C_j e^{2i\zeta_j^2 x} [M(x,\zeta_j)]_2}{\zeta - \zeta_j}.
\end{equation}
If we impose the symmetries (\ref{Msymmetry}), equation (\ref{M1decompose}) can be written as
$$\begin{pmatrix} \overline{M_{22}(x, \sigma \bar{\zeta})} \\ \overline{M_{12}(x, \sigma\bar{\zeta})} \end{pmatrix} = \begin{pmatrix} 1	\\	0 \end{pmatrix} + \sum_{j = 1}^{2N} \frac{C_j e^{2i\zeta_j^2 x}}{\zeta - \zeta_j}\begin{pmatrix} M_{12}(x, \zeta_j) \\ M_{22}(x, \zeta_j) \end{pmatrix}.$$
Evaluation at $\sigma\bar{\zeta}_k$ yields 
\begin{equation}\label{algebraicsystem}
\begin{pmatrix} \overline{M_{22}(x, \zeta_k)} \\ \overline{M_{12}(x, \zeta_k)} \end{pmatrix} = \begin{pmatrix} 1	\\	0 \end{pmatrix} + \sum_{j = 1}^{2N} \frac{C_j e^{2i\zeta_j^2 x}}{\sigma\bar{\zeta}_k - \zeta_j}\begin{pmatrix} M_{12}(x, \zeta_j) \\ M_{22}(x, \zeta_j) \end{pmatrix}, \qquad k =1, \dots, 2N.
\end{equation}
Solving this algebraic system for $M_{12}(x, \zeta_j)$, $M_{22}(x, \zeta_j)$, $j = 1, \dots, 2N$, and substituting the solution into (\ref{M1decompose}) yields an explicit expression for $[M(x,\zeta)]_1$. If we define $[M(x,\zeta)]_2$ by (\ref{Msymmetry}), the second residue condition (\ref{residue2}) is a consequence of symmetry. This solves the Riemann-Hilbert problem.
The potential $u_x(x)$ can be found by noticing that (\ref{M1decompose}) implies
$$M_{12}(x, \zeta) = \sigma \sum_{j = 1}^{2N} \bar{C}_j e^{-2i\bar{\zeta}_j^2 x} \overline{M_{22}(x,\zeta_j)} \frac{1}{\zeta} + O\left(\frac{1}{\zeta^2}\right), \qquad \zeta \to \infty.$$
Therefore, by (\ref{recoverqRH}),
\begin{equation}\label{recoversoliton} 
 u_x(x) = 2im(x)e^{4 \sigma i\int_{-\infty}^{x} |m|^2 dx'}, \qquad m(x) = \sigma \sum_{j = 1}^{2N} \bar{C}_j e^{-2i\bar{\zeta}_j^2 x} \overline{M_{22}(x,\zeta_j)}.
\end{equation}

\subsection{One-soliton solution}
In this section we derive explicit formulas for the one-soliton solutions. Assume $N=1$ so that there are two zeros of $a(\zeta)$: $\zeta_1$ in the first quadrant and $\zeta_2 = -\zeta_1$ in the third quadrant. Using that $C_1 = C_2$ we find that the algebraic system (\ref{algebraicsystem}) reduces to the following two equations:
\begin{align*} 
&\overline{M_{22}(x, \zeta_1)} = 1 + C_1 e^{2i\zeta_1^2 x}\left(\frac{1}{\sigma\bar{\zeta}_1 - \zeta_1} - \frac{1}{\sigma\bar{\zeta}_1 + \zeta_1}\right) M_{12}(x, \zeta_1)
	\\
&\overline{M_{12}(x, \zeta_1)} = C_1 e^{2i\zeta_1^2 x} \left(\frac{1}{\sigma\bar{\zeta}_1 - \zeta_1}+ \frac{1}{\sigma\bar{\zeta}_1 + \zeta_1}\right)M_{22}(x, \zeta_1).
\end{align*}
Solving for $\overline{M_{22}(x, \zeta_1)}$ we find
$$\overline{M_{22}(x, \zeta_1)} = \frac{1}{1  + \frac{4\sigma \zeta_1^2 |C_1|^2e^{2i(\zeta_1^2 - \bar{\zeta}_1^2)x}}{(\zeta_1^2 - \bar{\zeta}_1^2)^2}}$$

We parametrize the zero $\zeta_1$ and the normalization constant $C_1$ in terms of four parameters $\Delta > 0, \gamma \in (0, \pi), x_0 \in \R,$ and $\Sigma_0 \in \R$, according to
$$\zeta_1^2 = \Delta^2(-\sigma\cos \gamma + i\sin \gamma), \qquad C_1 = i \Delta (\sin \gamma) e^{4 \sigma i\int_{-\infty}^\infty |m|^2 dx'} e^{2i\Sigma_0}e^{2\Delta^2  x_0 \sin\gamma}.$$
Then, writing $\theta = \Delta^2 (x - x_0) \sin \gamma$, we find
$$\overline{M_{22}(x, \zeta_1)} = \frac{1}{1 + e^{-4\theta} e^{-\sigma i \gamma}}.$$
Furthermore, by (\ref{recoversoliton}),
$$m(x) = -2i \sigma \Delta (\sin \gamma) e^{-4 \sigma i\int_{-\infty}^\infty |m|^2 dx'} \frac{e^{-2i\Sigma + 2\theta}}{e^{4\theta} + e^{-\sigma i \gamma}},$$
where 
$$\Sigma = -\sigma \Delta^2 x \cos \gamma  + \Sigma_0.$$
Moreover, using
$$|m|^2 =  \frac{4\Delta^2 \sin^2 \gamma e^{4\theta}}{(e^{4\theta} + e^{\sigma i \gamma})(e^{4\theta} + e^{-\sigma i \gamma})},$$
we deduce that
$$e^{2 \sigma i\int_{-\infty}^{x} |m|^2 dx'} = e^{2 \sigma i\int_{-\infty}^\infty |m|^2 dx'}\frac{e^{4\theta} + e^{-\sigma i \gamma}}{e^{4\theta} + e^{\sigma i \gamma}}.$$
Substituting this into (\ref{recoversoliton}), we find
\begin{equation}\label{uxsoliton}  
  u_x = 4\sigma \Delta \sin \gamma e^{-2i\Sigma + 2\theta}\frac{e^{4\theta} + e^{-\sigma i \gamma}} {(e^{4\theta} + e^{\sigma i \gamma})^2}.
\end{equation}
Integration yields
$$u = -\frac{2i\sin\gamma }{\Delta}\frac{e^{- \sigma i \gamma}e^{-2i\Sigma}}{e^{2\theta} + e^{\sigma i \gamma}e^{-2\theta}}.$$
The time evolution is determined by the requirement $C_1(t) = C_1(0)e^{2i\eta_1^2 t}$:
\begin{align*}
& \Sigma_{0t} = \text{Re}\left(\eta_1^2\right) = -\alpha\beta-\sigma \alpha \Delta^2 \cos{\gamma}\left(1 + \frac{\beta^2}{4\Delta^4}\right), 
	\\
& x_{0t} = -\frac{\text{Im}\, \eta_1^2}{\Delta^2 \sin{\gamma}} = -\alpha \left(1 - \frac{\beta^2}{4\Delta^4}\right).
\end{align*}
The final expression for the one-soliton solutions of equation (\ref{GNLSgauge}) parametrized by the four parameters
\begin{equation}\label{4parameters}
  \gamma \in (0, \pi), \qquad \Delta > 0, \qquad \Sigma_0 \in\R, \qquad x_0 \in \R,
\end{equation}
is
\begin{equation}\label{usoliton}
u(x,t) =  -\frac{2i\sin \gamma}{\Delta}\frac{e^{- \sigma i \gamma}e^{-2i\Sigma(x,t)}}{e^{2\theta(x,t)} + e^{\sigma i \gamma}e^{-2\theta(x,t)}},
\end{equation}
where 
$$\Sigma(x,t) = -\alpha \beta t -\sigma\Delta^2 \cos \gamma\left( x + \alpha \left(1 + \frac{\beta ^2}{4\Delta^4}\right)t\right) + \Sigma_0$$
and
$$\theta(x,t) = \Delta^2 \sin \gamma \left(x - x_0 + \alpha \left(1 - \frac{\beta ^2}{4\Delta^4}\right)t\right).$$

\begin{remark}\upshape
The expression for $u_x$ as given in equation (\ref{uxsoliton}) coincides with formula (33c) in \cite{K-N} for the one-soliton solution of the derivative NLS equation. This is because the $x$-part of the Lax pair (\ref{lax}) coincides with the isospectral problem for the derivative NLS equation written in terms of $u_x$ \cite{K-N}.
\end{remark}

\subsection{Other kinds of soliton solutions}
We finally make some comments regarding the existence of other kinds of soliton solutions for equation (\ref{GNLS}). One well-known property of the CH equation is that it admits peaked solitons (these are continuous solutions with a peak at their crest) \cite{C-H}. Since (\ref{GNLS}) is related to the NLS equation by a procedure analogous to that which gives the Camassa-Holm equation from KdV, it is natural to ask whether (\ref{GNLS}) also exhibits weak solutions of this kind. However, a first analysis indicates that there are no natural candidates for peakon solutions of equation (\ref{GNLS}) cf. \cite{Lfiber}.

On the other hand, equation (\ref{GNLS}) does admit rational solitons, i.e. smooth solitons with algebraic decay at infinity. These solutions arise from the solitons with exponential decay of equation (\ref{usoliton}) in the limit $\gamma \uparrow \pi$. If $u(x,t)$ is given by (\ref{usoliton}), then
$$u(x,t) = u_r(x,t) + O(\pi - \gamma), \qquad \gamma \uparrow \pi,$$
where
$$u_r(x,t) = \frac{2 i e^{-2 i \sigma  \left(\Delta^2 x + \alpha  \left(\frac{\beta
   }{2 \Delta }- \sigma\Delta  \right)^2 t + \sigma  \Sigma_0 \right)} \Delta }{4 \Delta ^4 (x-x_0) -   \alpha  \left(\beta ^2-4 \Delta ^4\right) t + i \sigma  \Delta ^2}.$$
We claim that $u_r(x,t)$ is a rational soliton. Indeed, for the ranges of the parameters $\Delta, \Sigma_0, x_0$ specified in (\ref{4parameters}), $u_r(x,t)$ is a smooth function of $(x,t)$ with algebraic decay as $|x| \to \infty$, and it can be verified directly that it satisfies equation (\ref{GNLSgauge}).

\appendix
\section{Derivation of a Lax pair}
\renewcommand{\theequation}{A.\arabic{equation}}\nequation
In this appendix we briefly indicate how the bi-Hamiltonian structure was used to derive the Lax pair (\ref{lax}).

\begin{enumerate}
\item Consider the recursion operator $L = \theta_2 \theta_1^{-1}$ with adjoint $L^\dagger = \theta_1^{-1}\theta_2$. The bi-Hamiltonian theory implies that the equation
\begin{equation}\label{Ldaggerxpart}
L^\dagger \begin{pmatrix} \varphi_1 \\ \varphi_2 \end{pmatrix} = \lambda \begin{pmatrix} \varphi_1 \\ \varphi_2 \end{pmatrix},
\end{equation}
is the $x$-part of a Lax pair with spectral parameter $\lambda$ for (\ref{GNLS}) in terms of the `squared eigenfunctions' $\varphi_1(x,t)$ and $\varphi_2(x,t)$ cf. \cite{F-A}.

\item Assume the eigenfunctions $\psi_1(x,t)$ and $\psi_2(x,t)$ satisfy an $x$-part of the form
$$\begin{cases}
	& \psi_{1x} = \lambda h_1 \psi_1 + u_{12}\psi_2, \\
	& \psi_{2x} = u_{21}\psi_1 + \lambda h_2 \psi_2,
		\end{cases}$$
for some coefficients $h_1, h_2, u_{12}, u_{21}$. The `squared eigenfunctions' $\varphi_1 = \bar{\psi}_1\psi_2$ and $\varphi_2 = \bar{\psi}_2\psi_1$ then satisfy
\begin{align} \nonumber
	 \varphi_{1xx} + \biggl(\lambda(h_1& - h_2) - \frac{u_{21x}}{u_{21}}\biggr)\varphi_{1x} 
	 	\\ \label{varphixpart}
	& + \biggl(-2u_{12}u_{21} - (h_1 - h_2)\lambda \frac{u_{21x}}{u_{21}}\biggr)\varphi_1 + 2u_{21}^2 \varphi_2 = 0,
	\\  \nonumber
	 \varphi_{2xx} + \biggl(-\lambda(h_1 &- h_2) - \frac{u_{12x}}{u_{12}}\biggr)\varphi_{2x}
	 \\  \nonumber
	&  + \biggl(-2u_{12}u_{21} + (h_1 - h_2)\lambda \frac{u_{12x}}{u_{12}}\biggr)\varphi_2 + 2u_{12}^2 \varphi_1 = 0.
\end{align}
The coefficients $h_1, h_2, u_{12}, u_{21}$ can now be chosen so that (\ref{varphixpart}) coincides with (\ref{Ldaggerxpart}). After redefining $\lambda$ we find that the $x$-part takes the form
\begin{equation}\label{psialphabetaxpart}
\begin{cases}
	& \psi_{1x} = i(\alpha\lambda^2 - \beta)\psi_1 + \lambda q \psi_2, \\
	& \psi_{2x} = -i(\alpha\lambda^2 - \beta)\psi_2 + \lambda r \psi_1,
	\end{cases}
\end{equation}	
for some coefficients $\alpha$ and $\beta$ independent of $\lambda$.

\item Assume a $t$-part of the form
\begin{equation}\label{ABCDtpart}
\begin{cases}
	& \psi_{1t} = A\psi_1 + B\psi_2,	\\
	& \psi_{2t} = C \psi_1 - A\psi_2,
	\end{cases}
\end{equation}
and take the coefficients $A,B,C$ to be Laurent series in $\lambda$. In view of the analogous situation for the Camassa-Holm equation (\ref{CH}), one might expect $A,B,$ and $C$ to include some negative powers of $\lambda$. After equating terms at each order of $\lambda$, we find that
equation (\ref{GNLS}) is the compatibility condition between (\ref{psialphabetaxpart}) and (\ref{ABCDtpart}) provided that the following equations are satisfied:
$$A =\frac{i \sigma \gamma^2}{\nu^2} \lambda^2 + \frac{i \gamma}{\nu^2} + \frac{i \sigma}{2} u v +\frac{i \sigma}{4 \nu^2}\frac{1}{\lambda^2}, \quad B = \frac{\gamma}{\nu}q \lambda + \frac{\sigma}{2\nu}u \frac{1}{\lambda}, \quad C =  \frac{\gamma}{\nu}r \lambda + \frac{\sigma}{2\nu}v \frac{1}{\lambda},$$
and
$$\alpha =\frac{\sigma\gamma}{\nu}, \quad \beta = -\frac{1}{2 \nu}, \quad q = u + i\nu u_x, \quad r = v - i\nu v_x, \quad r = \sigma \bar{q}, \quad v = \sigma \bar{u}.$$

\item Transform the Lax pair (\ref{psialphabetaxpart})-(\ref{ABCDtpart}) by letting
\begin{equation}\label{UzetaVdef}
U =\frac{\nu}{\sqrt{\alpha}} \begin{pmatrix} 0	& -e^{-ix/\nu}u	\\
e^{ix/\nu} v	&	0 \end{pmatrix}, \qquad \zeta = -i\sqrt{\alpha}\lambda, \qquad V = \begin{pmatrix} e^{-\frac{ix}{2\nu}}\psi_1	\\
e^{\frac{ix}{2\nu}}\psi_2  \end{pmatrix}.
\end{equation}
Then equation (\ref{GNLS}) is the condition of compatibility for the two vector eigenvalue equations
\begin{equation}\label{newlax}
\begin{cases}
	& V_x + i\zeta^2 \sigma_3 V = \zeta U_x V,	\\
	& V_t + \frac{i\gamma}{\nu}\left(\zeta^2  - \frac{1}{\nu} + \frac{1}{4\nu^2 \zeta^2}\right)\sigma_3 V = \left(\frac{\zeta\gamma}{\nu} U_x +\frac{i\gamma}{2\nu^3}\sigma_3(-U^2 +\frac{1}{\zeta} U)\right)V.
	\end{cases}
\end{equation}	
These equations imply the Lax pair (\ref{lax}) after suitable redefinitions of $\alpha$ and $\beta$.

\end{enumerate}

 \bigskip
\noindent
{\bf Acknowledgement} {\it The authors acknowledge support from a Marie Curie Intra-European Fellowship and EPSRC.}

\bibliography{is}

\end{document}